# RASPBERRY PI AND ARDUINO UNO WORKING TOGETHER AS A BASIC METEOROLOGICAL STATION

José Rafael Cortés León[1] and Ricardo Francisco Martínez-González[2], Anilú Miranda Medinay[3] and Luis Alberto Peralta-Pelaez[3]

[1]Department of Electric-Electronic Engineering, Instituto Tecnologico de Veracruz (TecNM), Veracruz, Mexico
[2]Department of Electric-Electronic Engineering, Instituto Tecnologico de Veracruz (TecNM), Veracruz, Mexico
[3]Department of Chemistry-Biochemistry Engineering, Instituto Tecnologico de Veracruz (TecNM), Veracruz, Mexico

## ABSTRACT

*The present paper describes a novel Raspberry Pi and Arduino UNO architecture used as a meteorological station. One of the advantages of the proposed architecture is the huge quantity of sensors developed for its usage; practically one can find them for any application, and weather sensing is not an exception. The principle followed is to configure Raspberry as a collector for measures obtained from Arduino, transmitting occurs via USB; meanwhile, Raspberry broadcasts them via a web page. For such activity is possible thanks to Raspbian, a Linux-based operating system. It has a lot of libraries and resources available, among them Apache Web Server, that gives the possibility to host a web-page. On it, the user can observe temperature, humidity, solar radiance, and wind speed and direction. Information on the web-page is refreshed each five minute; however, measurements arrive at Raspberry every ten seconds. This low refreshment rate was determined because weather variables normally do not abruptly change. As an additional feature, system stores all information on the log file, this gives the possibility for future analysis and processing.*

## KEYWORDS

*Raspberry Pi, Arduino UNO, Meteorological station, Novel architecture.*

## I. INTRODUCTION

In Instituto Tecnologico de Veracruz, some professors have put their efforts together to produce energy by cleaner ways than fossil fuel combustion. Most common options are solar panels and wind turbines [1], both options have a strong dependency of weather emplacement conditions. The current project arises as a need to determine parameters during local studies for renewable energy production. And to give it solution, an embedded system was developed.

An embedded system is a computational system dedicated for a specific purpose [2][3][4]. Customers usually are in touch with them and do not notice them; almost every "smart" is a regular product with an embedded system, which possibilities internet connection, memory, interoperability or simply upgrades human interaction [5][6].

Often in order to reduce costs or energy consumption, embedded system hardware are specifically designed [7]; despite its design, they are often based on a microcontroller (μC) or microprocessor (μP). Where a μC is a single chip with μP, memory and peripherals included [8]; so, in one way or another, a microprocessor is always part of embedded systems.

An interesting fact of embedded systems is their variety, some of them just execute simple instructions; meanwhile, others support operating system (OS). No matter which hardware a developer uses, it usually can be programmed using C language or any other high-level language, allowing developers to select correct hardware and to get focus on system clocks as it.

There is an extended quantity of available hardware embedded systems for developers; on the current project, two were chosen: Raspberry Pi and Arduino UNO.

## 1.1 RASPBERRY PI

Eben Upton, Rob Mullins, Jack Lang and Alan Mycroft, at the United Kingdom, developed Raspberry Pi. It was created for promoting computer science teachings into elementary school students [9]. The Raspberry Pi is based on a Broadcom system-on-chip device; which had a 700 MHz processor, a graphics processor unit and 512 megabytes of random access memory (RAM) [10].

Nowadays, there are various models of Raspberry Pi but project works on model B. It has eight general-purpose input/outputs, two USB ports, a high definition media interface (HDMI) output, and other non-relevant features for this project. Additionally, Raspberry Pi needs an operating system which is stored on a secure digital (SD) card. Operating system selection was not easy, because of the good quantity of them; the more relevant one are Raspbian, Pidora, and RISC OS. Raspbian was selected for current project for the wide diversity of tutorials and info available online. Raspbian is a free-license Linux operating system based on Debian and optimized for its usage with Raspberry Pi hardware [11].

Raspberry plays as the central recollecting unit and it is good enough because of its advanced and sophisticated features; however, for sensing and working as collecting satellite unit it is oversized, so a smaller device is required. There are many options, but for the current project, the choice is Arduino UNO [12].

## 1.2 ARDUINO UNO

Arduino is an open-source physical computing platform based on a simple microcontroller board and a development environment for writing software for such board. Arduino started in 2005 as a project for students at the Interaction Design Institute Ivrea [13]. Today, Arduino platform has several configurations, however, according to the number of bits they use, they can be classified in two; the based on 8-bit Atmel AVR microcontrollers, and other ones that use 32-bit Atmel ARM processors [14]. Whatever the platform, it provides a set of digital and analog input/output pins that allows interacting with various extension boards.

There is an increasing number of Arduino boards; the vendor exposes that their selling finances project. The boards are designed to cover, practically, any application. For the current project, Arduino UNO was selected.

Once the hardware was selected, for programming, Arduino web-page released an integrated development environment (IDE). It is based on C and C++ programming languages, and probably its ace up the sleeve for Arduino, a software library is called "Wiring", and it makes simpler the communication with extension boards. Once hardware for the satellite units was defined, meteorological stations will be discussed in next section.

### 1.3 METEOROLOGICAL STATION

From ancient times, humankind has required knowing climate conditions. Knowledge gives the opportunity to prognosticate future conditions and elaborate forecasts. Using prognosticate, farmers could anticipate meteors; taking right decisions and saving crops and cattle [15]. In most recent times owing to global warming, forecasting has increased its importance, not just for crops but energy production; let us recoil, renewable energy has a high dependency of weather [16][17]. Before the digital era, weather stations were facilities with atmospheric measuring equipment; it used to be large, voluminous and manually operated. Nonetheless, everything changes and after digital era, such facilities have evolved, decreased their size, and obtained autonomy. Nowadays is possible to have the same sensibility from old fashion facilities in small and automatic boxes, not bigger than a little container for shoes.

### 1.4 LOCAL CONDITIONS

The project will work for determining specific weather conditions surrounding Veracruz city. According to the National Oceanic and Atmospheric Administration, Veracruz has a tropical coast climate. From June to October is wet and rainy, exceeding 25ºC. The dry season spans from November to May, with cooler temperatures and few precipitations. The average yearly precipitation is 1564 mm, being July the wettest month. June and August are the hottest months, and January is the coolest [18].

On next section, the parts of hardware system will be described, as well as its physical interconnection. The software is divided on two, one for Arduino routine and another for Raspberry configuration and programming, and they are in section three. The obtained results and its discussion appear on section four; meanwhile, conclusions are in section five.

## 2. HARDWARE SYSTEM DESCRIPTION

The system is formed by a Raspberry Pi Model B, it does not have enough USB ports for connect a keyboard, mouse, Wi-Fi dongle, and connect Arduino; therefore, an expansion board was utilized. Expansion board gives 2 additional USB ports, it eliminates the need for a HDMI to VGA converter cable due to it has a converter included and a VGA output. It also has a real-time clock and supplies power to model B.

As it was mentioned a Wi-Fi dongle is used, it possibilities connection to wireless local area networks (WLAN) and internet access. The Wi-Fi dongle model is TP-LINK TL-WN722N, this dongle is approved by Raspberry Pi community for its compatibility with Raspbian.

For installing and user usage, a keyboard and mouse are needed, they are regular USB ones. A flat-screen monitor allows watching OS graphical interface. To connect Arduino UNO, a USB type A to B cable is used; Arduino oversees sensors management, complete configuration appears in Figure 1.

Raspberry connections are in blue and red ones are Arduino connections. Arduino relates to sensors, according to needs; solar luminescence, humidity, wind speed, and temperature. Once architecture was built up, next step is operating system loading for Raspberry Pi.

International Journal of Computer Science & Information Technology (IJCSIT) Vol 9, No 5, October 2017

## 3. PROGRAMMING

After Raspbian was loaded, Arduino IDE for Linux is next; because of Raspbian are based on Linux, this is possible. Having the complete system, it works as one from the beginning, getting easier architecture debugging process.

Programming was divided into three parts: the program for Arduino UNO, another for Raspberry Pi and web-page creation.

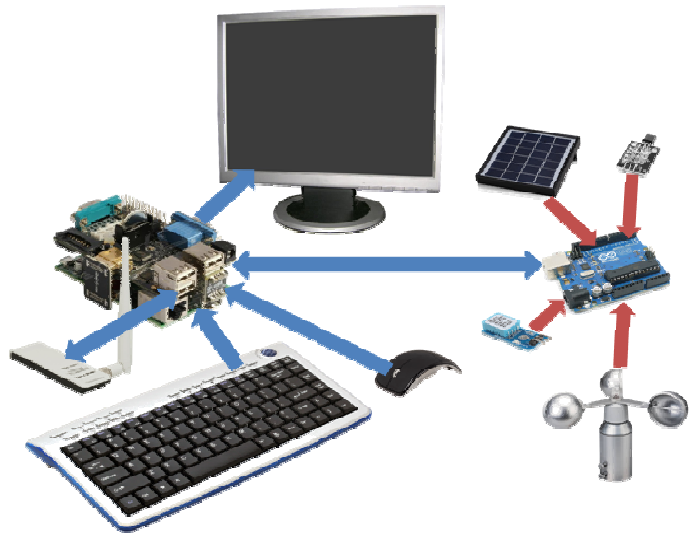

**Figure 1**. Planned hardware configuration for meteorological station

### 3.1 PROGRAM FOR ARDUINO UNO

For Arduino programming, its platform utilizes a language like C. Program firstly defines which pin is used for what sensor. In a second place, it configures serial communication, and finally, it asks sensors for their measurements. Flow diagram for the program is depicted in Figure 2.

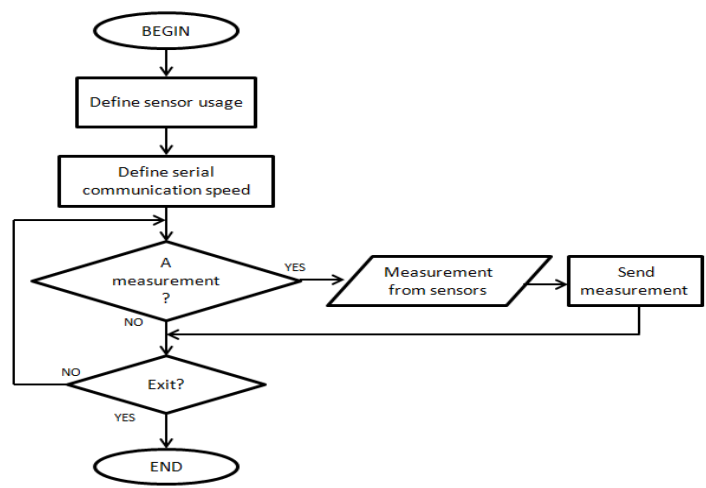

Figure 2. Flow diagram for Arduino program

100

A good aspect for Arduino is its interoperability; if for any reason another Arduino is used, the program still works with no change.

### 3.2 PYTHON PROGRAM FOR RASPBERRY PI

Some persons affirm that Pi from Raspberry Pi comes after Python. As it was explained at first section; Raspberry was created for teaching, and as Python is a simple language, makes it ideal for such purpose.

The program for Raspberry Pi starts firstly defining serial connection speed; for a second, it asks Arduino for measures. The final step is to print them on screen and in a log file; complete flow diagram for the program is in Figure 3.

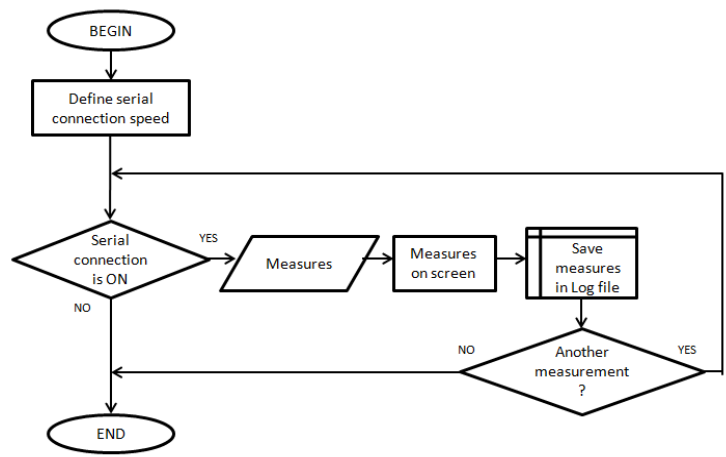

**Figure 3.** Flow diagram for Raspberry Pi program

Both programs enable communication between Raspberry Pi and Arduino; however, spread measurements, for it a web-page is used.

### 3.3 WEB SERVER AND WEBPAGE CREATION

Linux has a web server available named Apache, it is the most widely-used web server software in the world. It also requires PHP complements for web-page creation and PostgreSQL for database management.

Another Python program collects data measurements from the log file and modifies HTML code contented in web-page; after it, the user must refresh browser for displaying new measurement.

### 4. RESULTS AND DISCUSSION

At first, the communication inter-system was set to maximum baud rate, around 1 megabaud per second, but transmission had some errors, in consequence, lower baud rate was determined. The fact does not represent a problem for system efficiency since weather changes are typically slow. Another advantage of reducing baud rate is energy consumption decrement, even though such expenditure was not quantified. 9600 baud rate was the selected one, it guarantees a free-error transmission; although, in real scenarios, transmission rate can be reduced more During Arduino and Raspberry tests, measurements were displayed on the console; the update times were four-five seconds, mostly because response time for sensors. Another characteristic was low change rate, the prototype was tested under a controlled environment, thus sensors do not experiment abrupt measures.

Web-page was a complete success, maybe it is simple but it works, and it is loaded on Raspberry Pi. This fact possibilities later updates and modifications. The web-page is shown in Figure 4.

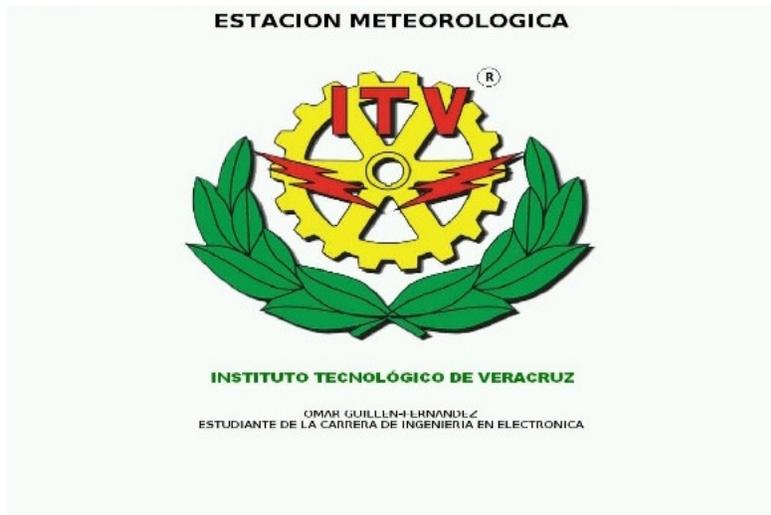

**Figure 4**. Meteorological station web-page

Raspberry Pi and Arduino UNO architecture, shown in Figure 5, was a keystone for web server creation. At supporting an operating system, a wide variety of options were available. The architecture also measures weather variable; its small size makes portability possible. And finally, using generic programs it is upgradeable. If better boards are released; the web server, python programs, and even Arduino programs can be reloaded at new ones.

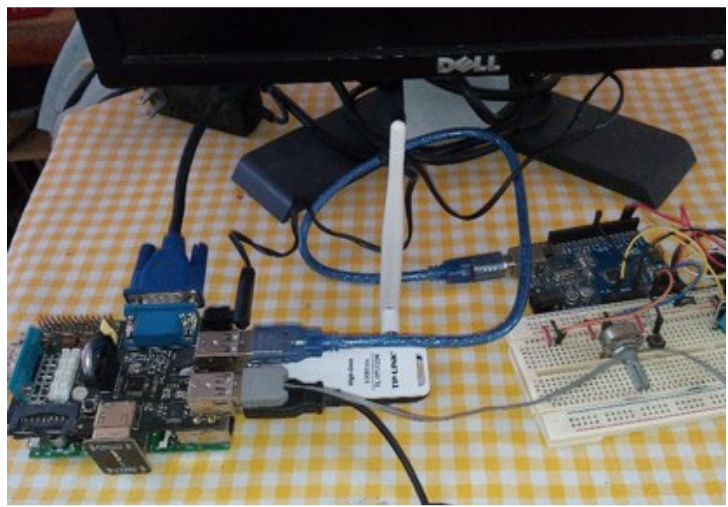

Figure 5. Raspberry Pi and Arduino UNO architecture

## 5. CONCLUSIONS

Some developing boards are so powerful that they can be loaded with an operating system. OS empowers several options, like a web-page or database support. The Raspberry Pi was developed for teachings, but its good design and support have increased its limits beyond its designers thought.

Using Arduino was simpler than we expected, there are libraries for almost everything. Libraries reduce programming troubles, increasing the time for other project activities. In recent times, the number of sensor boards designed for Arduino has grown up in unthinkable ways; there are practically sensors for anything, and more important making new ones is relatively easy.

Finally, architecture exceeds our initial expectations. It was designed as a meteorological station, and it measures weather variables and shows them on a web-page, which is hosted on same Raspberry, and refreshed every five minutes. It is also capable to save information into log files for post-processing. With no hesitations is possible to conclude that project was satisfying developed, thanks for the widely available information and resources for both devices integrated on proposed architecture.

## 6. FUTURE WORK

This was a basic station, we think to expand the number of measuring stations; in other words, to have a good number of Arduino board taking measures and wirelessly transmitting them to a base station, where a Raspberry Pi uploads them.

In addition, basic sensors were used; now that architecture was approved, to buy or to make more responsive and precise sensors is necessary. Although architecture was probed and approved, it can be upgraded. There is another more powerful Arduino, it supports an operating system and it has all classic Arduino features. If it works, we could save a Raspberry Pi board, installing a web server on this Arduino and taking measures on very same board.

And finally, web-page needs some improvements; create a more useful web-page, use graphs for weather data, have control access, elaborate a file with measurements, enable remote control station, and acquire a public domain.

## ACKNOWLEDGMENTS

Thanks for the financial support provided by Tecnológico Nacional de Mexicothrough the project: "Evaluación de la Generación de Bio-Electricidad, Depuración de Agua y Disminución de GEI como Servicios Ecositémicos que Proporcionan Humedales Artificiales", with number: 9hh5h1 (1980).

International Journal of Computer Science & Information Technology (IJCSIT) Vol 9, No 5, October 2017